\documentclass[sigconf, natbib=false]{acmart}

\usepackage{array}
\usepackage{caption}  
\usepackage{hyperref}

\AtBeginDocument{%
  }

\setcopyright{acmlicensed}          
\copyrightyear{2026}
\acmYear{2026}
\setcopyright{cc}
\setcctype{by}
\acmConference[DAC '26]{63rd ACM/IEEE Design Automation Conference}{July 26--29, 2026}{Long Beach, CA, USA}
\acmBooktitle{63rd ACM/IEEE Design Automation Conference (DAC '26), July 26--29, 2026, Long Beach, CA, USA}
\acmDOI{10.1145/3770743.3804197}
\acmISBN{979-8-4007-2254-7/2026/07}

\RequirePackage[
  datamodel=acmdatamodel,
  style=acmnumeric,
  ]{biblatex}

\makeatletter
\def\@authors{%
  \def\And{\unskip, }
  \def\AND{\unskip, }
  \def\LastAnd{, and }
  \mbox{\normalsize\bfseries\@authorlist}
}
\makeatother

\begin{document}

\title{Algorithm and Hardware Co-Design for Efficient Complex-Valued Uncertainty Estimation}





\author{Zehuan Zhang$^{1}$, Mark Chen$^{1}$, He Li$^{2}$, Wayne Luk$^{1}$}

\affiliation{%
  \institution{$^{1}$Department of Computing, Imperial College London \quad $^{2}$Southeast University}
  \country{}
}

\renewcommand{\shortauthors}{Zhang et al.}

\begin{abstract}
Complex-Valued Neural Networks (CVNNs) have significant advantages in handling tasks that involve complex numbers.
However, existing CVNNs are unable to quantify predictive uncertainty.
We propose, for the first time, dropout-based Bayesian Complex-Valued Neural Networks (BayesCVNNs) to enable uncertainty quantification for complex-valued applications, exhibiting broad applicability and 
efficiency for hardware implementation due to modularity.
Furthermore, as the dual-part nature of complex values significantly broadens the design space and enables novel configurations based on \textbf{layer-mixing} and \textbf{part-mixing}, we introduce an automated search approach to effectively identify optimal configurations for both real and imaginary components.
To facilitate deployment, we present a framework that generates customized FPGA-based accelerators for BayesCVNNs, leveraging a set of optimized building blocks.
Experiments demonstrate the best configuration can be effectively found via the automated search, attaining higher performance with lower hardware costs compared with manually crafted models. 
The optimized accelerators achieve approximately $4.5\times$ and $13\times$ speedups on different models with less than $10\%$ power consumption compared to GPU implementations, and outperform existing work in both algorithm and hardware aspects.
Our code is publicly available at: https://github.com/zehuanzhang/BayesCVNN.git.

\end{abstract}


\maketitle

\renewcommand{\footnoterule}{} 

\begingroup
\renewcommand{\thefootnote}{}  
\footnotetext{\vspace{-0.3mm}\footnotesize 
This work is supported in part by UK EPSRC (EP/X036006/1, EP/P010040/1,
EP/V028251/1, EP/S030069/1, UKRI256), in part by the National Natural Science Foundation of China under Grant 62304037, in part by the Natural Science Foundation of Jiangsu Province under Grant BK20230828, and in part by AMD.
\vspace{-0.3mm}}
\addtocounter{footnote}{-1}    
\endgroup

\vspace{-2mm}
\section{Introduction}

Deep neural networks have delivered extraordinary performance across various domains~\cite{ sharifani2023machine}. 
In practical scenarios such as radar imaging~\cite{mu2021cv, ouabi2020stochastic, cao2019pixel, gao2018enhanced}, weather forecasting~\cite{goh2006complex}, and magnetic resonance imaging~\cite{cole2021analysis, li2024classification, xiao2022partial}, data are collected as complex values containing both real parts and imaginary parts.
To handle complex-valued tasks effectively, Complex-Valued Neural Networks (CVNNs)~\cite{popa2017complex, trabelsi2017deep} have been introduced,
and have shown significant advantages compared with real-valued neural networks (RVNNs) in complex-valued tasks~\cite{barrachina2021complex}. 
Despite their potential, existing CVNNs lack the ability for uncertainty estimation, which is crucial for reliability and trustworthiness. Uncertainty estimation prevents potential risks especially for health-related and safety-critical applications~\cite{zou2023review, zhang2024accelerating}, 
since it enables assessment of prediction confidence, highlighting processes that require human attention.


In real domains, Bayesian Neural Networks (BayesNNs)~\cite{neal1992bayesian} have
emerged to endow models capable of estimating uncertainty.
Previous work~\cite{jospin2022hands, goan2020bayesian} has explored efficient approaches to perform inference. However, these efforts focused on real domains, leaving complex domains underexplored. 
Directly employing RVNNs to process complex-valued data can result in intrinsic loss of information in imaginary parts~\cite{lv2019hybrid, sunaga2019land, wang2020deepcomplexmri}.
Therefore, to further enhance the reliability of CVNNs, there is an urgent need to develop Bayesian Complex-Valued Neural Networks (BayesCVNNs), extending Bayesian approximations to complex domains.

Nevertheless, to design and deploy BayesCVNNs for practical applications, several key challenges exist.
\textbf{First}, although real-domain Bayesian approximation methods~\cite{blundell2015weight, gal2016dropout} significantly reduce operational complexity, their effectiveness in complex domains remains unexplored. 
\textbf{Second}, the inherent dual-component property of complex values leads to a larger model design space.
It is a heuristic-driven process to manually design model architectures, making it particularly challenging to develop optimal model designs
that satisfy diverse demands of various applications, 
especially considering the impact of real and imaginary parts.
\textbf{Third}, dual components and complex-domain arithmetic~\cite{lee2023exploiting} drastically increase memory and computational costs. 
Representing a single complex number with two real-valued elements doubles the memory footprint, and introduces at least a fourfold increase in computational complexity.
These incurred costs hinder the practical deployment of BayesCVNNs.

To address the \textbf{first} challenge, this paper extends Monte-Carlo Dropout~\cite{gal2016dropout} from real to complex domains, providing mathematical analysis that dropout enables Bayesian approximations for CVNNs. 
The proposed methodology utilizes dropout layers as plug-and-play modules, which exhibit broad applicability across mainstream models.
The dropout operations can be implemented by independent and modular engines, facilitating hardware implementation.
To address the \textbf{second} challenge, we first conduct a quantitative analysis, which indicates that BayesCVNNs exhibit an exponentially
broader design space with respect to the number of layers (Section~\ref{sec:3_2}), compared with dropout-based BayesNNs in real domains.
To effectively identify the optimal configurations, we construct a design space that enables \textbf{layer-mixing} and \textbf{part-mixing} configurations (Section~\ref{sec:4_1}). We further propose an evolutionary search approach considering hardware constraints to automatically identify the optimal model designs that outperform manually crafted models with higher algorithmic performance and lower hardware costs.
To address the \textbf{third} challenge, we develop efficient FPGA-based hardware modules by leveraging complex-valued operational characteristics, and design a set of fundamental building blocks for complex-valued layers to generate specialized hardware accelerators.
Specifically, complex-valued dropout layers contain switches to allow part-mixing configurations. Furthermore, considering application demands and hardware resource constraints, latency-opt and resource-opt hardware mapping schemes are provided. 
The generated accelerators, with uncertainty estimation capabilities,
outperform GPU implementations and existing accelerators in speed and power.
Moreover, ablation studies examine the hardware costs associated with RVNN-to-CVNN and CVNN-to-BayesCVNN conversions, quantifying FPGA deployment overheads.

Our contributions can be summarized as follows.


\vspace{-0.5mm}
\begin{itemize}
    \item A novel dropout-based BayesCVNN that enables reliable uncertainty estimation for complex-valued applications. 
    This complex-valued dropout module supports plug-and-play deployment in CVNNs, showing algorithmic generality and hardware advantages.

    \item An automated search approach to effectively find the dropout configurations for real and imaginary parts that meet both algorithmic targets and hardware constraints. This approach enables layer-mixing and part-mixing configurations, identifying the optimal configuration within an exponentially larger design space than conventional BayesNNs.
    \item An algorithm-hardware co-design framework to generate customized FPGA accelerators for BayesCVNNs based on a set of building blocks with hardware mapping schemes.

\end{itemize}

\vspace{-2mm}
\section{Background and Related Work}\label{sec:II}
\vspace{-1mm}
\subsection{Bayesian Neural Network}

Normal neural networks represent model parameters as point values.
Unlike those, BayesNNs treat weights as probability distributions. 
The learning process is based on Bayesian theory:
\vspace{-1mm}
\begin{equation}
P(\mathbf{w}|\mathcal{D}) = \frac{P(\mathcal{D}|\mathbf{w})P(\mathbf{w})}{P(\mathcal{D})}
\end{equation}
where $\mathbf{w}$ represents weight parameters and $\mathcal{D}$ means input data. $P(\mathbf{w})$ is the prior probability distribution, which represents the probability of model parameters prior to observing any data.
$P(\mathcal{D}|\mathbf{w})$ is the likelihood, representing the probability of data $\mathcal{D}$ given model parameters $\mathbf{w}$.
$P(\mathcal{D})$ is called evidence, which is given by: 

\vspace{-4mm}
\begin{equation}
P(\mathcal{D}) = \int P(\mathcal{D}|\mathbf{w}) P(\mathbf{w}) d\mathbf{w}.
\end{equation}
\vspace{-4mm}


Due to the high complexity of neural networks, analytical posterior distributions are computationally intractable. To address this, various approximation methods have been proposed~\cite{neal2011mcmc,welling2011bayesian}, among which Monte Carlo Dropout~\cite{gal2016dropout} gains attention for its efficiency. Dropout can be used for Bayesian approximation: during inference, BayesNNs perform multiple forward passes for sampling, yielding a set of outputs, where the average is the final prediction and the standard deviation measures associated uncertainty.



\vspace{-3mm}
\subsection{Complex-Valued Neural Network}

Complex numbers include real and imaginary parts.
CVNN layer operations can be performed via executing real-valued operations following complex-domain arithmetic~\cite{lee2023exploiting}. 
For instance, in complex convolutional layer operations, the data are denoted as $A_R + A_Ii$ and the filters are denoted as $W_R + W_Ii$.  To calculate the results, four sub-operations are needed, i.e. $W_R*A_R$, $W_R*A_I$, $W_I*A_R$, $W_I*A_I$. Afterwards, addition and subtraction are performed to produce final outputs. Specially, $(W_R*A_R - W_I*A_I)$ represents the real part, and $(W_R*A_I + W_I*A_R)$ represents the imaginary part of outputs. 
Complex fully connected layers are processed in similar ways. 
Regarding other types of CVNN layers, including pooling and activation layers, real-valued operations only need to be applied separately to real and imaginary parts to produce outputs.

\vspace{-3mm}
\subsection{Related Work}

\textbf{Dropout-based BayesNN:}
Prior work~\cite{gal2016dropout, ferianc2021combinet, zhang2019confidence,  kendall2015bayesian} has demonstrated that dropout-based BayesNNs can provide high-quality uncertainty estimation, thereby enhancing model trustworthiness. Nevertheless, no studies have extended dropout-based approximations to complex domains. 

\noindent
\textbf{Domain-Specific Accelerators:}
Research has been conducted on accelerators for BayesNNs~\cite{zhang2024hardware, fan2023monte, fan2022fpga, cai2018vibnn, fan2021high, ferianc2021optimizing, awano2020bynqnet, fujiwara2021asbnn, fan2022accelerating, que2025trustworthy}.
The intrinsic characteristics of BayesNNs and the exhibited sparsity are exploited to attain high hardware performance. 
Furthermore, accelerators for CVNNs have appeared recently. The studies~\cite{lee2023exploiting,peng2021binary} proposed quantization-level optimizations and developed hardware accelerators for CVNNs. Work~\cite{ahmad2024fpga} explored polar-form operations to optimize CVNN accelerators, boosting speed and power efficiency.
However, existing accelerators are unable to support BayesCVNNs.

This paper proposes a framework to develop dropout-based BayesCVNNs and specialized accelerators, exhibiting high generality. An evolutionary search is employed to find the optimal model configuration. Additionally, a set of fundamental hardware modules are introduced to produce FPGA hardware accelerators.

\vspace{-3mm}
\section{Model Design Based on Dropout}

\vspace{-1mm}
\subsection{Mathematical Analysis}

Exactly computing the posterior distribution over weights given the data is computationally intractable for neural networks. 
Variational inference~\cite{blei2017variational} approximates the true posterior with a simpler distribution. 
~\cite{gal2016dropout} demonstrates applying dropout is equivalent to sampling from a Bernoulli distribution over weights during each forward pass. 
Consequently, training with dropout can be interpreted as minimizing a variational objective, with KL divergence as a key component, thereby serving as a Bayesian approximation.
The derivation process consists of four steps: 
(\textit{S1})~framing dropout as a variational distribution, 
(\textit{S2})~deriving the training objective: minimizing the KL divergence to optimize the variational distribution, 
(\textit{S3})~approximating the training objective
(\textit{S4})~obtaining model uncertainty with stochastic forward passes.

To demonstrate applying dropout for CVNNs is a mathematically valid Bayesian approximation, we extend the sampling processes to CVNNs (corresponding to (\textit{S1})) and provide the variational objectives (corresponding to (\textit{S2})). 
The remaining steps follow the approach in~\cite{gal2016dropout}.
In CVNN, complex weights are represented as:

\vspace{-5mm}
\begin{equation}
   W = W_{\text{real}} + jW_{\text{imag}} 
\end{equation}
where \( W_{\text{real}} \) and \( W_{\text{imag}} \) are the real and imaginary parts of the complex weights. 
The dropout layers can be inserted into both parts, denoted as \textbf{Scenario1}, or into either part, denoted as \textbf{Scenario2}.
In \textbf{Scenario2}, we assume dropout layers are applied only to the real parts. The analysis is similar if they are applied to the imaginary parts. 
\textbf{Note that Scenario2 highlights a distinctive optimisation opportunity in the complex domain, enabling different mixed configurations involving real and imaginary parts.}

To approximate the posterior \( p(W_{\text{real}}, W_{\text{imag}} | X,Y) \),  a tractable distribution \( q( W_{\text{real}}, W_{\text{imag}} ) \) is defined as Equation~\ref{eq4} in \textbf{Scenario1} and Equation~\ref{eq5} in \textbf{Scenario2}.
\vspace{-0.5mm}
\begin{align}
W_{\text{real}} &= M_{\text{real}} \cdot \text{diag}(z_{\text{real}}), & W_{\text{imag}} &= M_{\text{imag}} \cdot \text{diag}(z_{\text{imag}}) \label{eq4} \\
W_{\text{real}} &= M_{\text{real}} \cdot \text{diag}(z_{\text{real}}), & W_{\text{imag}} &= M_{\text{imag}} \label{eq5}
\end{align}
\vspace{-0.5mm}
where \( M_{\text{real}} \) and \( M_{\text{imag}} \) are deterministic parameters, and \( z_{\text{real}}, z_{\text{imag}} \) are Bernoulli random variables representing the dropout masks. 
This distribution over the weights serves as the variational approximation to the true posterior distribution over complex weights.

Similar to the real-valued case, the training objective in complex-valued neural networks is to minimize the divergence between the posterior and the defined tractable distribution. This can be written as Equation~\ref{eq6} in \textbf{Scenario1}:
\vspace{-1.5mm}
\begin{equation}
\begin{aligned}
\int q(W_{\text{real}}, W_{\text{imag}}) \left[ - \log p(Y | X, W_{\text{real}}, W_{\text{imag}}) \right] dW_{\text{real}} dW_{\text{imag}} \\
+ \text{KL}(q(W_{\text{real}}, W_{\text{imag}}) \| p(W_{\text{real}}, W_{\text{imag}}))
\label{eq6}
\end{aligned}
\end{equation}

\vspace{-1.5mm}
\noindent and Equation~\ref{eq7} in \textbf{Scenario2}:
\vspace{-2mm}
\begin{equation}
\begin{aligned}
\int q(W_{\text{real}}) \left[ - \log p(Y | X, W_{\text{real}}, M_{\text{imag}}) \right] dW_{\text{real}} \\
+ \text{KL}(q(W_{\text{real}}) || p(W_{\text{real}}))
+ \text{KL}(M_{\text{imag}} || p(W_{\text{imag}}))
\label{eq7}
\end{aligned}
\end{equation}
where the \(\text{KL}(\cdot, \cdot)\) measures the similarity between the posterior distribution of the model weights and the defined weights, which are sampled following Bernoulli distributions. 
By applying the subsequent steps ((\textit{S3}) \& (\textit{S4})) outlined in~\cite{gal2016dropout} to Equations~\ref{eq6} and \ref{eq7}, it can derive the dropout-based approximations on both the real and imaginary components for uncertainty prediction.




This objective retains the key characteristics of Bayesian inference by treating the complex weights as stochastic variables. 
Randomness is introduced even if dropout is only applied to the real part, enabling the model to capture and model uncertainty. This stochastic treatment allows the model to account for variability in its predictions, thereby enabling uncertainty estimation.

\vspace{-2mm}
\subsection{Model Analysis}\label{sec:3_2}

The mathematical analysis shows dropout enables Bayesian approximations for complex-domain models, enabling the construction of dropout-based BayesCVNNs with the following advantages.

\textbf{Algorithm Advantages:}
Dropout is a technique broadly utilized in modern models, which can be incorporated following both convolutional layers and fully connected layers. Therefore, most existing CVNNs can be converted to BayesCVNNs by adding dropout layers
as intermediate layers, demonstrating broad applicability.

\textbf{Hardware Advantages:}
In dropout-based BayesCVNNs, dropout layers function as independent and modular components that can be inserted into the model with minor structural modifications. This modularity benefits hardware design, as only the dropout engines are required and can be seamlessly integrated into existing hardware architectures, enhancing productivity and portability.

\vspace{-2mm}
\subsection{Design Space}\label{sec:3_2}

To construct dropout-based BayesCVNNs, the specified dropout layers are utilized to perform Bayesian sampling, referred to as Bayesian layers. Due to the inherent characteristics of complex values, dropout-based BayesCVNNs allow for multiple configurations, resulting in a significantly larger design space compared with BayesNNs in real domains.

\begin{figure}[htbp]
\vspace{-3pt}
\centerline{\includegraphics[width=3.0in]{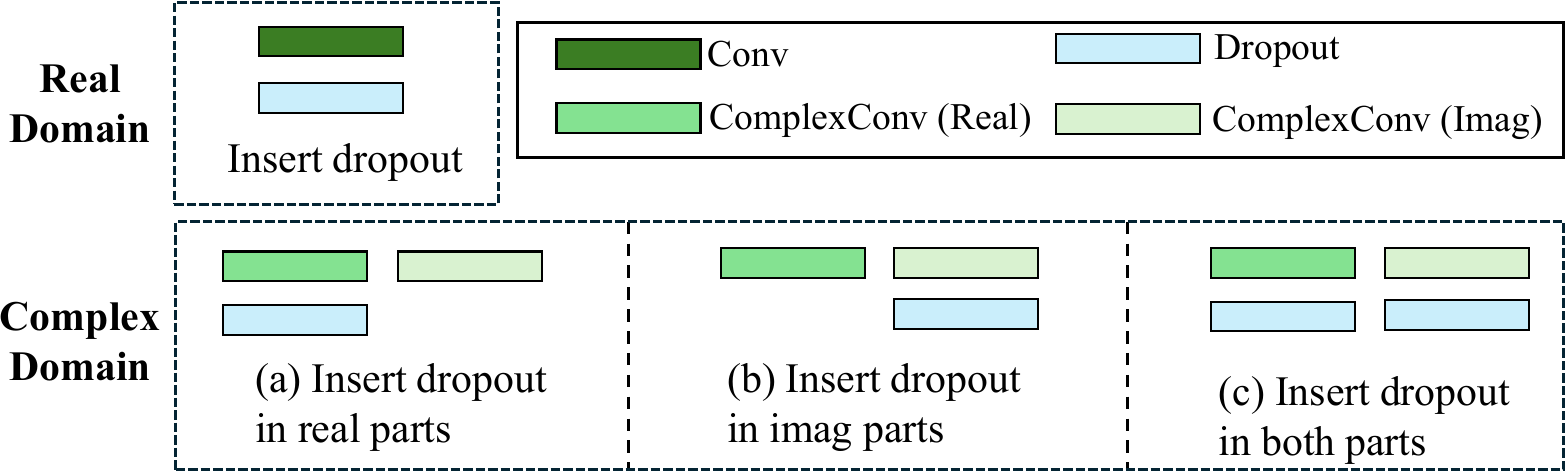}}
\captionsetup{skip=1mm} 
\caption{ Bayesian Layer Configurations. }
\label{fig:3}
\vspace{-20pt}
\end{figure}


\vspace{-1mm}
Let $N$ be the number of Bayesian layers. In each Bayesian layer, configurations differ as shown in Fig.~\ref{fig:3}. Three configurations are allowed in complex domains, leading to a total of $3^N$ configurations. By contrast, in real domains, the Bayesian layer configuration is fixed.
Therefore, BayesCVNNs exhibit a $3^N$ times larger design space than BayesNNs in real domains, offering flexibility to explore trade-offs between accuracy, uncertainty, and hardware costs.


\vspace{-2mm}
\section{Framework for Accelerator Generation}

\begin{figure*}[htbp]
\centerline{\includegraphics[width=6.4in]{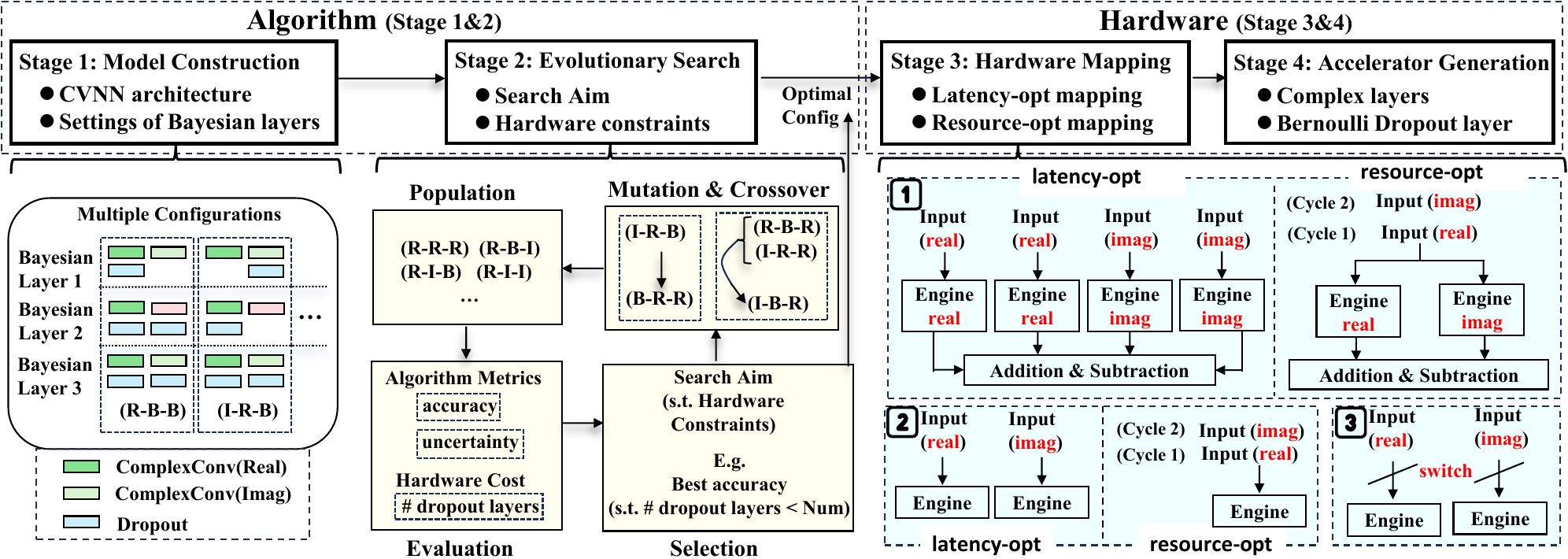}}
\captionsetup{skip=1mm}
\caption{ Overview of the Framework. }
\label{fig:4_1}
\vspace{-4mm}
\end{figure*}

A framework is developed to produce the optimal model configuration and generate customized hardware accelerators, as shown in Fig.$~\ref{fig:4_1}$. It consists of four stages: (1) model construction, (2) evolutionary search, (3) hardware mapping, (4) accelerator generation.


\vspace{-3mm}
\subsection{Model Construction}\label{sec:4_1}

Stage 1 involves the construction of BayesCVNNs and model training.
To build dropout-based BayesCVNNs, the original CVNN model architecture, the position and number of Bayesian layers, and the dropout rates are given. 
A set of BayesCVNN models are constructed through inserting Bernoulli dropout layers in different ways, constituting the design space. Each Bayesian layer allows for three configurations (i.e. inserting dropout in real/imaginary/both parts) and layer configurations can differ as well, as shown in the example in Fig.$~\ref{fig:4_1}$.
As a result, the design space includes \textbf{layer-mixing} and \textbf{part-mixing} configurations, harnessing the full potential of varied models. 
Afterwards, models are trained on the datasets.


\vspace{-3mm}
\subsection{Evolutionary Search}
As previously analyzed, the design of BayesCVNNs enables a $3^N$ larger design space. Algorithmic performance dramatically varies, and hardware costs, which are mainly decided by the number of dropout layers (demonstrated in Section~\ref{sec:V_C}), are not consistent among diverse network configurations.
To find the optimal configuration efficiently, an evolutionary search accounting for hardware constraints is conducted, comprising four steps: (a) population, (b) evaluation, (c) selection and (d) mutation$\&$crossover.

During the \textit{population} step, the initial pool of configurations is randomly generated to a predefined size. Each is represented by a combination of dropout designs in multiple layers.
The \textit{evaluation} step measures both algorithm performance and hardware costs, including accuracy and the quality of uncertainty estimation and the number of dropout layers.
In the \textit{selection} step, the top-performing models while satisfying hardware constraints are selected as the pool of parents for \textit{mutation and crossover}.
In the fourth step, a portion of the selected parent configurations undergo
mutation while the remaining configurations undergo crossover.
\textit{Mutation} means each configuration is changed randomly, while \textit{crossover} randomly performs layer-wise mixture of Bayesian layers between two configurations to produce a new configuration. 
The mutation and crossover stage conducts layer-wise exploration, allowing for the update of \textbf{layer-mixing} and \textbf{part-mixing} configurations.
As a result, new model configurations are produced to update the population pools.
The four steps are repeated several iterations to produce the final model configuration best achieving the search objective and meeting hardware constraints.

\vspace{-3mm}
\subsection{Hardware Mapping}

Existing Qkeras libraries do not support complex layers. To enable the construction of BayesCVNNs, we develop a set of building blocks with hardware mapping schemes. According to operational characteristics, complex layers can be categorized into three classes.

The \textit{first class} includes complex convolutional and complex fully connected layers, both requiring four sub-operations along with extra addition and subtraction operations. The weight reusability exists among the sub-operations, enabling hardware resource reuse. Therefore, we propose two design schemes: latency-opt and resource-opt schemes as illustrated in Fig.$~\ref{fig:4_1}$.
The latency-opt scheme deploys four separate operational engines for each sub-operation, allowing for parallel processing and effectively reducing latency at the cost of higher resource consumption. In contrast, the resource-opt scheme deploys one real engine and one imaginary engine, processing real and imaginary parts of input data sequentially, which exploits the weight reusability to save resource utilization but increases latency due to the reduced level of parallelism.

The \textit{second class} includes complex pooling and complex activation layers, which simply apply real-domain functions to real and imaginary parts without extra calculations. Therefore, the engines can also be reused. As shown in Fig.$~\ref{fig:4_1}$, the latency-opt scheme deploys two engines for parallel processing of both parts of input data, while the single engine is shared in the resource-opt scheme, resulting in a trade-off between latency and resource consumption.

The \textit{third class} is the Bayesian layer, the computation of which varies depending on the specific layer configurations. Naively placing dropout layers after both the real and imaginary parts is infeasible, as only one part requires dropout operations in mixed configurations. To support all three possible configurations within each Bayesian layer, we incorporate switches for dropout engines as shown in Fig.$~\ref{fig:4_1}$, each dedicated to handling one part of input data. Adjusting switch settings ensures functional correctness of the target configurations while minimizing resource consumption.

\vspace{-5mm}
\subsection{Accelerator Generation}

Hardware accelerators are generated based on our customized design templates. The generated HLS-based BayesCVNN accelerators can then be processed by Vivado HLS for synthesis and implementation, producing the final bitstream for onboard testing.

\begin{figure}[htbp]
\centerline{\includegraphics[width=2.4in]{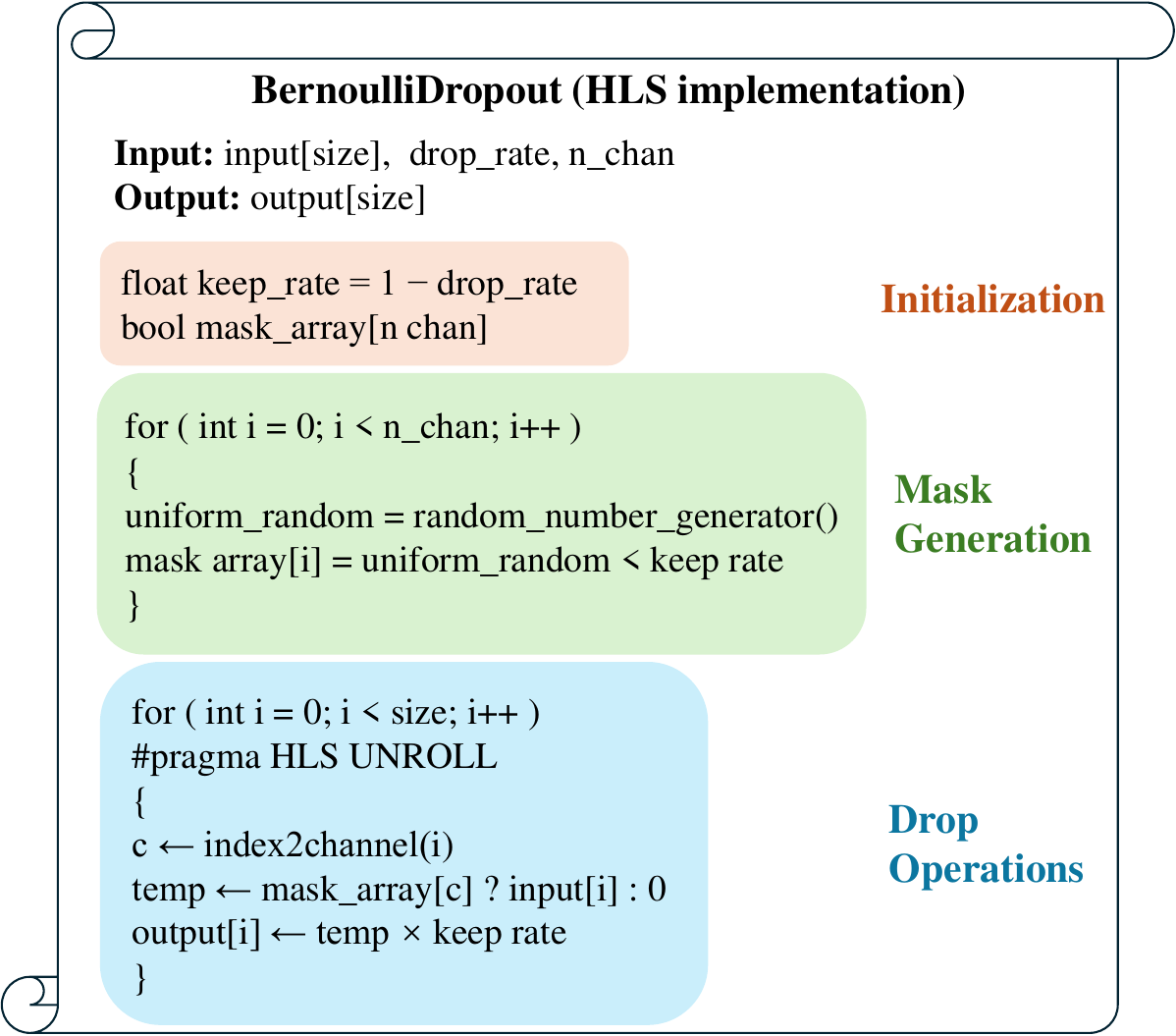}}
\captionsetup{skip=1mm}
\caption{ Pseudocode of Bernoulli Dropout. }
\label{fig:4_4}
\end{figure}

The pseudocode of HLS-based implementation of Bernoulli dropout is presented in Fig.~\ref{fig:4_4}. It drops entire channels of feature maps. During the mask generation process, the $mask\_array$ is generated based on $keep\_rate$, the size of which equals to the number of channels. When iterating the $mask\_array$, each point is indexed to identify which channel it belongs to through the $index2channel$ function, and dropping operations are implemented.

\vspace{-4mm}
\vspace{-5mm}

\begin{table*}[htbp]
\captionsetup{skip=1pt} 
\caption{Results on different datasets (R, I, and B represent inserting dropout layers into real, imaginary and both parts).} 
\label{tab:5_1} 
\centering 

\begin{tabular}
{
>{\centering\arraybackslash}p{1.5cm}
|p{0.1cm}p{0.1cm}>{\centering\arraybackslash}p{0.9cm}
|p{0.1cm}p{0.1cm}>{\centering\arraybackslash}p{0.9cm}
|p{0.8cm}p{0.8cm}p{0.8cm}
|p{0.8cm}p{0.8cm}>{\centering\arraybackslash}p{0.8cm} 
}
\hline

\textbf{Model} & \multicolumn{3}{c|}{Bayesian ComplexLeNet} & \multicolumn{3}{c|}{CHAR} & \multicolumn{3}{c|}{CASR-small} & \multicolumn{3}{c}{CASR-large} \\ \hline
\textbf{Dataset} & \multicolumn{3}{c|}{MNIST} & \multicolumn{3}{c|}{HAR} & \multicolumn{3}{c|}{Chicago} & \multicolumn{3}{c}{Houston} \\ \hline

\textbf{Config} 
& \multicolumn{1}{c|}{\begin{tabular}[c]{@{}c@{}}Manual\\ (B-B-B)\end{tabular}} 
& \multicolumn{1}{c|}{\begin{tabular}[c]{@{}c@{}}MaxAcc\\ (I-I-B)\end{tabular}} 
& \begin{tabular}[c]{@{}c@{}}MinECE\\ (R-B-R)\end{tabular} 
& \multicolumn{1}{c|}{\begin{tabular}[c]{@{}c@{}}Manual\\ (B-B-\\B-B-B)\end{tabular}} 
& \multicolumn{1}{c|}{\begin{tabular}[c]{@{}c@{}}MaxAcc\\ (R-I-\\I-I-R)\end{tabular}} 
& \begin{tabular}[c]{@{}c@{}}MinECE\\ (B-R-\\B-B-B)\end{tabular} 
& \multicolumn{1}{c|}{\begin{tabular}[c]{@{}c@{}}Manual\\ (B-B-B)\end{tabular}} 
& \multicolumn{1}{c|}{\begin{tabular}[c]{@{}c@{}}MaxAcc\\ (I-R-I)\end{tabular}} 
& \multicolumn{1}{c|}{\begin{tabular}[c]{@{}c@{}}MinECE\\ (R-B-R)\end{tabular}} 
& \multicolumn{1}{c|}{\begin{tabular}[c]{@{}c@{}}Manual\\(B-B-B-\\B-B-B)\end{tabular}} 
& \multicolumn{1}{c|}{\begin{tabular}[c]{@{}c@{}}MaxAcc\\ (R-R-R-\\R-B-R)\end{tabular}} 
& \begin{tabular}[c]{@{}c@{}}MinECE\\ (B-I-I-\\I-B-R)\end{tabular} \\ \hline

\textbf{Accuracy(\%)}      
& \multicolumn{1}{c|}{98.78} & \multicolumn{1}{c|}{\textbf{99.12}} & 99.07   
& \multicolumn{1}{c|}{82.81} & \multicolumn{1}{c|}{\textbf{89.54}} & 77.06        
& \multicolumn{1}{c|}{79.91} & \multicolumn{1}{c|}{\textbf{81.84}} & \multicolumn{1}{c|}{80.85}       
& \multicolumn{1}{c|}{85.34} & \multicolumn{1}{c|}{\textbf{87.10}} & 86.54           
\\ \hline
\textbf{ECE(\%)}           
& \multicolumn{1}{c|}{0.18} & \multicolumn{1}{c|}{0.16} & \textbf{0.06}                   
& \multicolumn{1}{c|}{0.59} & \multicolumn{1}{c|}{0.65} & \textbf{0.53}             
& \multicolumn{1}{c|}{1.06} & \multicolumn{1}{c|}{1.38} & \multicolumn{1}{c|}{\textbf{0.77}}        
& \multicolumn{1}{c|}{6.17} & \multicolumn{1}{c|}{7.43} & \textbf{1.32}                                  
\\ \hline
\textbf{\#Dropout}           
& \multicolumn{1}{c|}{6} & \multicolumn{1}{c|}{4} & 4                   
& \multicolumn{1}{c|}{10} & \multicolumn{1}{c|}{5} & 9             
& \multicolumn{1}{c|}{6} & \multicolumn{1}{c|}{3} & \multicolumn{1}{c|}{4}        
& \multicolumn{1}{c|}{12} & \multicolumn{1}{c|}{7} & 8                                  
\\ \hline
\end{tabular}
\vspace{-3mm}
\end{table*}

\section{Experiments and Evaluation}

The proposed BayesCVNNs are implemented using PyTorch 1.9.0.
Vivado-HLS 2019.2 is used for hardware implementation. QKeras 0.9.0 is used for quantization. The produced C-synthesis reports provide the performance of latency and resource consumption. The place-and-route optimizations are implemented by Vivado 2019.2 for the final designs. The FPGA board is Xilinx Kintex XCKU115.

\vspace{-3mm}
\subsection{Effectiveness of Evolutionary Search}
\vspace{-1.5pt}

The proposed BayesCVNNs are evaluated on both real-valued and complex-valued datasets.
We construct a complex-version Bayesian LeNet for MNIST~\cite{1571417126193283840} classification.
For complex-valued datasets, referring to~\cite{yang2022radar}, we build the CHAR model with five complex convolutional layers and select the radar-based human activity recognition (HAR) dataset~\cite{yang2022radar} for evaluations.
Moreover, we use the S1SLC$\_$CVDL datasets~\cite{asiyabi2023s1slc},
which comprise complex-valued synthetic aperture radar data patches from different scenes for classification. Two subsets are sampled from Chicago and Houston scenes. 
To assess scalability, we developed two BayesCVNN models—CSAR-small and CSAR-large—consisting of five and ten complex-valued convolutional layers, respectively. These models are evaluated on the Chicago and Houston scenes.
The numbers of specified Bayesian layers in Bayesian ComplexLeNet, CHAR, CSAR-small, and CSAR-large are set as three, five, three and six.
The sampling number is three.
Regarding the search settings, the population size is eight and the portion for mutation is $50\%$. The probability of mutation and crossover is 0.5.
More details are available in our codes.

\vspace{-1.5pt}
Experimental results including model configurations, algorithmic results and hardware costs are presented in Table~\ref{tab:5_1}. The algorithmic metrics include the accuracy and the expected calibration error (ECE)~\cite{guo2017calibration}, which assesses the quality of uncertainty estimation. Lower ECE indicates higher uncertainty estimation quality.
The total number of dropout layers is calculated as hardware costs. 
The configuration with dropout applied to both parts is referred to as the manual design. We adopt the highest accuracy and lowest ECE as search aims.
As shown in Table~\ref{tab:5_1}, the configurations found through search outperform manual designs.
Notably, all the searched configurations are mixed, bringing higher accuracy ($0.5\% \sim 2\%$) or better calibration capabilities than manually-designed uniform configurations.
\textbf{This highlights the significance of the development of layer-mixing and part-mixing design space and search, which enables the exploitation of diverse configurations.}
In addition, note that \textbf{the algorithmic performance improvements are not consistent with an increase in hardware costs.} 

\begin{figure}[htbp]
\centerline{\includegraphics[width=3.4in]{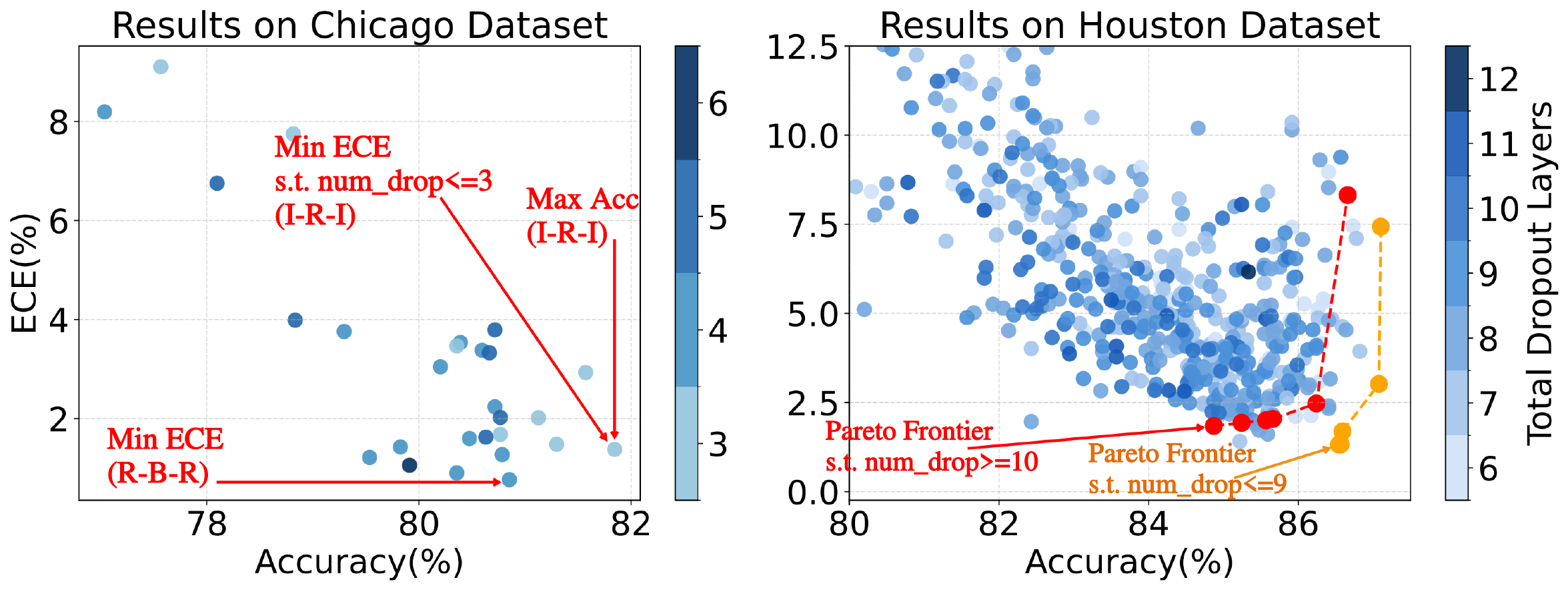}}
\captionsetup{skip=1mm}
\caption{ Search Results of Complex-Valued Tasks.}
\label{fig:5_1}
\end{figure}

To further demonstrate the effectiveness, 
we incorporate the hardware constraints in the search aims. We search for the lowest ECE with fewer than three dropout layers on the Chicago dataset. 
To compare in a straightforward way, we iterated through and measured the algorithmic performance of all the configurations along with hardware costs as shown in Fig.~\ref{fig:5_1} (left).
The identified configuration contains fewer dropout layers with minimal performance degradation, with ECE increasing by $0.61\%$.
Moreover, on the Houston dataset, we adjust search aims via assigning different weights to accuracy and uncertainty, while imposing hardware constraints that require the total number of dropout layers to be greater than 10 or fewer than 9, respectively, to identify Pareto-optimal designs that outperform any other designs in either accuracy or uncertainty.
The designs obtained from search with two different hardware constraints are highlighted in Fig.~\ref{fig:5_1} (right).
All the searched points form the reference Pareto frontiers, confirming the effectiveness of the automated search in identifying optimal solutions.
Notably, the searched designs on the Pareto frontier with fewer than 9 dropout layers exhibit superior algorithmic performance, further indicating that \textbf{increased hardware cost does not necessarily translate to improved algorithmic performance, underscoring the necessity of the evolutionary search.}
For real-world applications, model configurations can be obtained depending on specific optimization objectives, providing strong flexibility.

\vspace{-3mm}
\vspace{-3mm}

\subsection{Analysis of Mapping Schemes}

\begin{figure}[htbp]
\centerline{\includegraphics[width=3.4in]{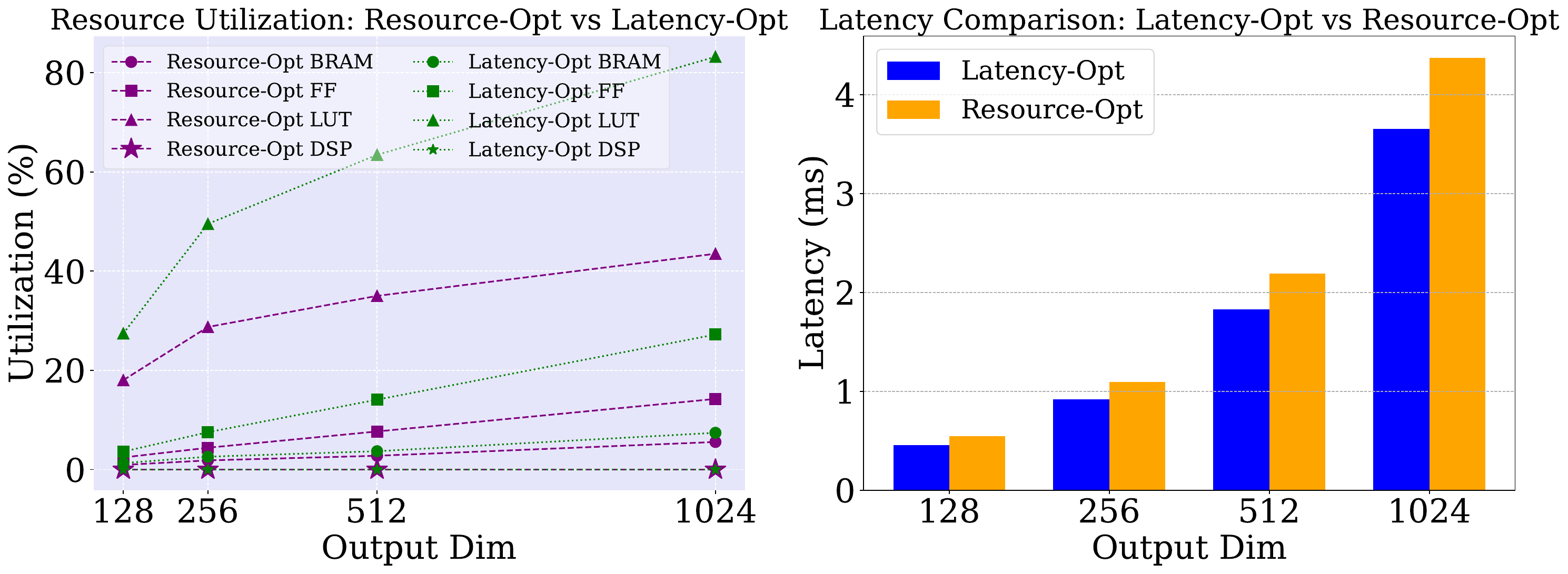}}
\vspace{-4mm}
\caption{ Resource and Latency of Mapping Schemes.}
\label{fig:5_2a}
\end{figure}

The performance difference between the two mapping schemes comes from the exploited reusability.
To quantitatively investigate the effects, we take complex fully connected layers as an example. We generate and synthesize the core operations, i.e. four sub-operations, with varying input and output dimensions. The input dimension is set as 128, while the output dimensions are set as 128, 256, 512, and 1024. Results are shown in Fig.~\ref{fig:5_2a}.

\vspace{-3mm}
\vspace{-3mm}

Four resource types (BRAM, DSP, FF, and LUT) are presented. 
As the output dimension increases, DSP utilization does not vary significantly, while the utilization of other resources shows an increasing trend. 
The latency-opt mapping scheme demands higher resource consumption than the resource-opt mapping scheme due to the higher parallelism.
In addition, latency comparison results exhibit that, as the output dimension grows, latency increases in both schemes, and the latency gap between the two schemes widens. 
That is because in the resource-opt mapping scheme, two parts of input data are processed sequentially by the shared operational engine, resulting in longer latency.


\subsection{Hardware Costs of Conversions}\label{sec:V_C}

The ablation study investigates the hardware costs associated with converting RVNNs to CVNNs, and CVNNs to BayesCVNNs.
The overheads of the former arise from real-to-complex layer conversion, while the latter's overheads stem from the incorporation of dropout layers.
For quantitative investigation, we implement LeNet5, ComplexLeNet5, and Bayesian ComplexLeNet5 with one Bayesian layer following the first complex convolutional layer, enabling three configurations. Since on-chip resources are sufficient to support all the models, latency-opt mapping schemes are adopted. 


\vspace{-1mm}
\begin{figure}[htbp]
\centerline{\includegraphics[width=3.1in]{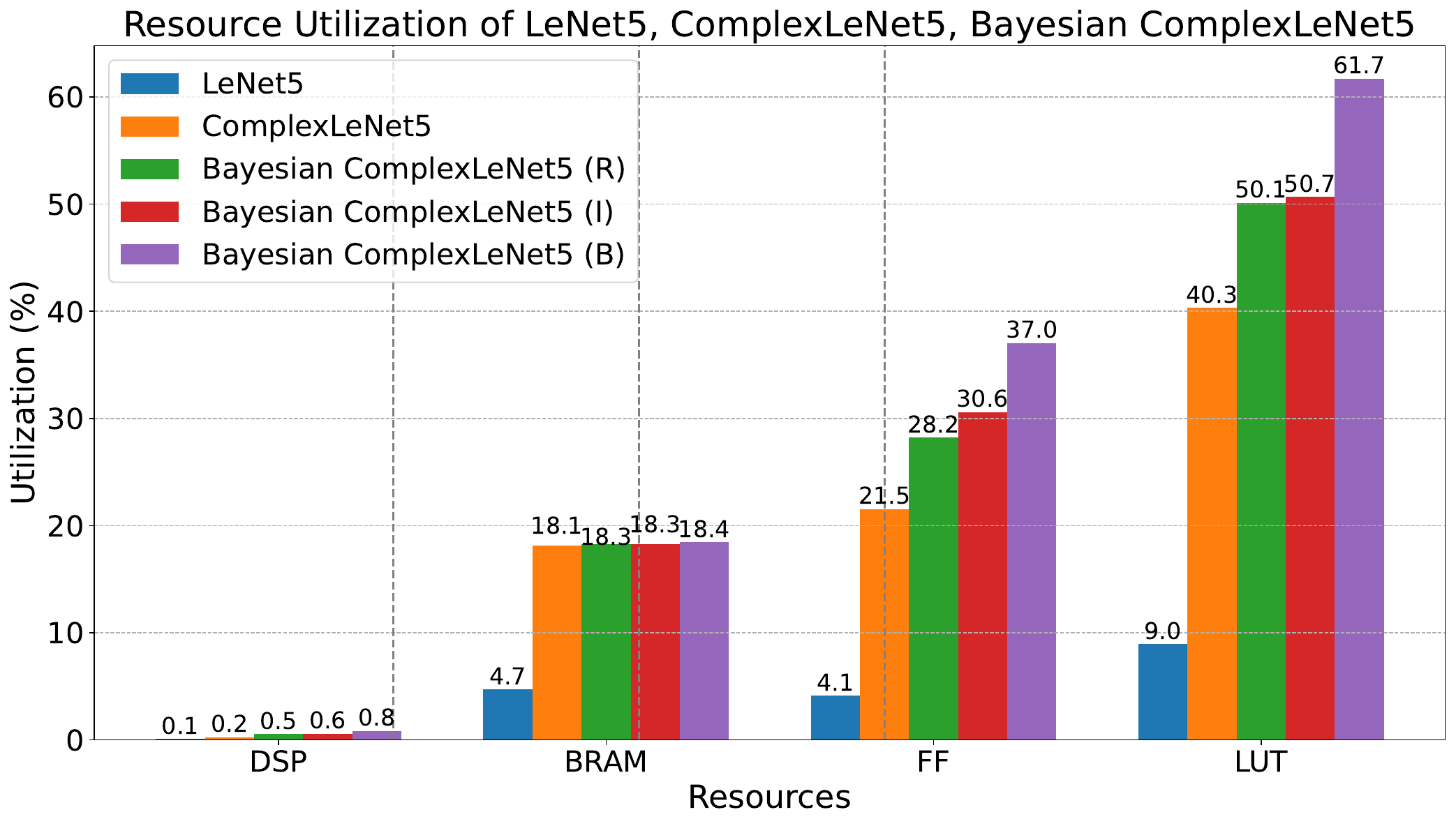}}
\captionsetup{skip=1mm}
\caption{ Resource Utilization of Different Models.}

\label{fig:5_5}
\end{figure}
\vspace{-2mm}

  


Fig.~\ref{fig:5_5} presents resource utilization of five generated accelerators.
\textbf{RVNN-to-CVNN conversions incur higher costs,} entailing a larger computational load, more complex control logic, and higher memory consumption due to the dual components of model parameters and complex-valued layer operations.
Comparing the results of LeNet5 and ComplexLeNet5 reveals that the conversion to complex layers increases DSP by $0.1\%$, BRAM by $13.4\%$, FF by $17.4\%$, and LUT by $31.3\%$.
Regarding the conversion for Bayesian inference, BRAM and DSP resource utilization exhibits a trivial increase while FF and LUT resource utilization requires relatively notable increases. 
That is due to the fact that dropout engines implement random dropping operations.
The configurations with one dropout layer, i.e. Bayesian ComplexLeNet5 (R) and Bayesian ComplexLeNet5 (I), show approximately consistent increases in resource utilization. Both require about $10\%$ increase in FF and LUT.
The Bayesian ComplexLeNet5 (B) containing two dropout layers demands more hardware costs, with FF resource utilization increasing by $15.5\%$ and LUT by $21.4\%$. 
This shows the hardware costs and the number of dropout layers exhibit a consistent growth trend, indicating that \textbf{employing the number of dropout layers to represent hardware costs is reasonable.}

\subsection{Comparisons with GPU and Related Work}

Table~\ref{tab:5_3} compares our FPGA accelerator with latency-opt mapping schemes against GPU implementation. To demonstrate the scalability, CSAR-small and -large models are implemented. 
In terms of the latency of implementing three Bayesian samplings, our FPGA designs achieve nearly $13\times$ and $4.5\times$ speed improvement for small and large models, respectively, compared to the GPU implementations,
despite its more advanced 8nm technology.
Furthermore, our designs demonstrate more than $90\%$ lower power consumption than the GPU, and over $100\times$ higher energy efficiency, consuming far less power for inference.
These gains can be attributed to the customized building blocks and latency-opt mapping schemes enabling high parallelism, delivering optimized solutions for dropout-based BayesCVNNs while demonstrating design scalability.

\begin{table}[hbtp]
\vspace{-5pt}
\centering
\captionsetup{skip=1pt}
\caption{FPGA design vs. GPU implementation performance.}
\begin{tabular}{|>{\raggedright\arraybackslash}p{2.4cm}|>{\centering\arraybackslash}p{1.0cm}|>{\centering\arraybackslash}p{1.0cm}|
                  >{\centering\arraybackslash}p{0.8cm}|>{\centering\arraybackslash}p{0.8cm}|}
\hline
   & \multicolumn{2}{c|}{GPU}    & \multicolumn{2}{c|}{Ours}    \\ \hline
\textbf{Platform}      & \multicolumn{2}{c|}{GeForce RTX 3090 Ti} & \multicolumn{2}{c|}{Kintex XCKU115} \\ \hline
\textbf{Frequency}    & \multicolumn{2}{c|}{1.40 GHz}     & \multicolumn{2}{c|}{181 MHz}      \\ \hline
\textbf{Technology}   & \multicolumn{2}{c|}{8nm}          & \multicolumn{2}{c|}{20nm}         \\ \hline
\textbf{Model}        & small         & large             & small            & large         \\ \hline
\textbf{Power(W)}     & 112             & 112                 & 3.81                & 4.65             \\ \hline
\textbf{\begin{tabular}[c]{@{}l@{}}Energy Costs\\ (mJ/forward pass)\end{tabular}} 
                      & 360.08             & 624.74                 & 0.89               & 5.64             \\ \hline
\textbf{Latency(ms)}  & 9.65             & 16.73                 & 0.70                & 3.64             \\ \hline
\end{tabular}
\label{tab:5_3}
\vspace{-5pt}
\end{table}

Until now no accelerators for BayesCVNNs have been
proposed. ~\cite{lee2023exploiting,peng2021binary} applied quantization-level
techniques to CVNNs.
Since their techniques are orthogonal
to ours, we do not compare and will explore
such optimizations for BayesCVNNs as future work.
Our design is compared with the CVNN accelerator in~\cite{ahmad2024fpga} and the traditional BayesNN accelerator in~\cite{fan2022accelerating,  fan2023monte, zhang2024hardware} as shown in Table~\ref{tab:5_4}.

To speed up inference,~\cite{ahmad2024fpga} transformed MNIST images to polar form representation but suffers from noticeable accuracy loss. 
We found utilizing smaller networks without polar-form conversions can achieve superior results, and implemented shallower Bayesian ComplexLeNet with fewer layers in rectangular form, outperforming~\cite{ahmad2024fpga} in both algorithm and hardware metrics. 
Using the same platform settings for a fair comparison, our design achieves higher accuracy, faster speed, lower power consumption, and uncertainty estimation capability. 
The work~\cite{zhang2024hardware} proposes search methods for conventional BayesNNs. Compared with that, our search approach leverages unique properties of complex values and considers \textbf{part-mixing} configurations, enabling exploration of a broader design space with \textbf{finer-grained} control.
Moreover, our search method integrates hardware constraints in a more \textbf{efficient} way by explicitly controlling the number of dropout layers, rather than by a complicated regression-based predictor as in~\cite{zhang2024hardware}.
Comparisons with~\cite{fan2022accelerating,  fan2023monte, zhang2024hardware} show that our design outperforms them in both latency and power consumption, even though our design supports arithmetic with complex numbers.

\begin{table}[hbtp]
\vspace{-5pt}
\centering
\captionsetup{skip=1pt}
\caption{Comparisons with related work.}
\begin{tabular}{|c|ccc|c|c|}
\hline

\textbf{}                                                    & \multicolumn{1}{l|}{\cite{fan2022accelerating}}  
& \multicolumn{1}{l|}{\cite{fan2023monte}}      
& {\cite{zhang2024hardware}}  
&    \cite{ahmad2024fpga}
& Ours       \\ \hline

Model Type                                                              & \multicolumn{3}{l|}{BayesNN}                                              & CVNN       & BayesCVNN  \\ \hline
Accuracy                                                                & \multicolumn{1}{l|}{-}     & \multicolumn{1}{l|}{\textbf{-}} & \textbf{-} & 88.3\%     & 95.4\%     \\ \hline
Latency(ms)                                                             & \multicolumn{1}{l|}{0.32}  & \multicolumn{1}{l|}{0.89}       & 0.91       & 1.60       & 0.27       \\ \hline
Uncertainty                                                            & \multicolumn{1}{l|}{$\checkmark$}  & \multicolumn{1}{l|}{$\checkmark$}       & $\checkmark$       & \textbf{$\times$}       & $\checkmark$      \\ \hline
Power(W)                                                                & \multicolumn{1}{l|}{43.6}  & \multicolumn{1}{l|}{4.6}        & 3.9       & 2.9       & 1.3       \\ \hline
\end{tabular}
\label{tab:5_4}
\vspace{-5pt}
\end{table}
\section{Conclusion}

This paper presents novel dropout-based BayesCVNNs and an automated search for identifying mixed configurations achieving higher performance at lower hardware costs than manually designed models.
To facilitate real-life deployment, an algorithm-hardware co-design framework is developed to generate customized FPGA-based accelerators via a set of building blocks with hardware mapping schemes.
Experiments demonstrate the generated accelerators outperform GPU implementations and existing accelerators.

\newpage
\printbibliography

@misc{asiyabi2023s1slc,
  author       = {H. N. Reza Mohammadi Asiyabi and M. Datcu and A. Anghel},
  title        = {S1SLC\_CVDL: A complex-valued annotated single look complex Sentinel-1 SAR dataset for complex-valued deep networks},
  howpublished = {\emph{IEEE Dataport}},
  note         = {Early access},
  year         = {2023},
  month        = apr,
  doi          = {10.1109/TGRS.2023.3267185}
}

@inproceedings{que2025trustworthy,
  title={Trustworthy Deep Learning Acceleration with Customizable Design Flow Automation},
  author={Que, Zhiqiang and Fan, Hongxiang and Figueiredo, Gabriel and Guo, Ce and Luk, Wayne and Yasudo, Ryota and Motomura, Masato},
  booktitle={Proceedings of the 15th International Symposium on Highly Efficient Accelerators and Reconfigurable Technologies},
  pages={1--13},
  year={2025}
}

@inproceedings{yang2022radar,
  title={Radar-based human activities classification with complex-valued neural networks},
  author={Yang, Ximei and Guendel, Ronny G and Yarovoy, Alexander and Fioranelli, Francesco},
  booktitle={2022 IEEE radar conference (radarConf22)},
  pages={1--6},
  year={2022},
  organization={IEEE}
}

@article{fan2022accelerating,
  title={Accelerating Bayesian neural networks via algorithmic and hardware optimizations},
  author={Fan, Hongxiang and Ferianc, Martin and Que, Zhiqiang and Niu, Xinyu and Rodrigues, Miguel and Luk, Wayne},
  journal={IEEE Transactions on Parallel and Distributed Systems},
  volume={33},
  number={12},
  pages={3387--3399},
  year={2022},
  publisher={IEEE}
}

@inproceedings{ferianc2021combinet,
  title={ComBiNet: Compact convolutional Bayesian neural network for image segmentation},
  author={Ferianc, Martin and Manocha, Divyansh and Fan, Hongxiang and Rodrigues, Miguel},
  booktitle={International Conference on Artificial Neural Networks},
  pages={483--494},
  year={2021},
  organization={Springer}
}

@article{zhang2019confidence,
  title={Confidence calibration for convolutional neural networks using structured dropout},
  author={Zhang, Zhilu and Dalca, Adrian V and Sabuncu, Mert R},
  journal={arXiv preprint arXiv:1906.09551},
  year={2019}
}

@article{kendall2015bayesian,
  title={Bayesian segnet: Model uncertainty in deep convolutional encoder-decoder architectures for scene understanding},
  author={Kendall, Alex and Badrinarayanan, Vijay and Cipolla, Roberto},
  journal={arXiv preprint arXiv:1511.02680},
  year={2015}
}

@article{sharifani2023machine,
  title={Machine learning and deep learning: A review of methods and applications},
  author={Sharifani, Koosha and Amini, Mahyar},
  journal={World Information Technology and Engineering Journal},
  volume={10},
  number={07},
  pages={3897--3904},
  year={2023}
}

@article{zou2023review,
  title={A review of uncertainty estimation and its application in medical imaging},
  author={Zou, Ke and Chen, Zhihao and Yuan, Xuedong and Shen, Xiaojing and Wang, Meng and Fu, Huazhu},
  journal={Meta-Radiology},
  pages={100003},
  year={2023},
  publisher={Elsevier}
}

@article{goan2020bayesian,
  title={Bayesian neural networks: An introduction and survey},
  author={Goan, Ethan and Fookes, Clinton},
  journal={Case Studies in Applied Bayesian Data Science: CIRM Jean-Morlet Chair, Fall 2018},
  pages={45--87},
  year={2020},
  publisher={Springer}
}

@article{goh2006complex,
  title={Complex-valued forecasting of wind profile},
  author={Goh, Su Lee and Chen, Mo and Popovi{\'c}, DH and Aihara, K and Obradovic, D and Mandic, DP},
  journal={Renewable Energy},
  volume={31},
  number={11},
  pages={1733--1750},
  year={2006},
  publisher={Elsevier}
}

@article{lv2019hybrid,
  title={A hybrid deep convolutional and recurrent neural network for complex activity recognition using multimodal sensors},
  author={Lv, Mingqi and Xu, Wei and Chen, Tieming},
  journal={Neurocomputing},
  volume={362},
  pages={33--40},
  year={2019},
  publisher={Elsevier}
}

@article{neal2011mcmc,
  title={MCMC using Hamiltonian dynamics},
  author={Neal, Radford M and others},
  journal={Handbook of Markov Chain Monte Carlo},
  volume={2},
  number={11},
  pages={2},
  year={2011},
  publisher={Chapman and Hall/CRC}
}

@inproceedings{welling2011bayesian,
  title={Bayesian learning via stochastic gradient Langevin dynamics},
  author={Welling, Max and Teh, Yee W},
  booktitle={Proceedings of the 28th international conference on machine learning (ICML-11)},
  pages={681--688},
  year={2011},
  organization={Citeseer}
}

@article{sunaga2019land,
  title={Land form classification and similar land-shape discovery by using complex-valued convolutional neural networks},
  author={Sunaga, Yuki and Natsuaki, Ryo and Hirose, Akira},
  journal={IEEE Transactions on Geoscience and Remote Sensing},
  volume={57},
  number={10},
  pages={7907--7917},
  year={2019},
  publisher={IEEE}
}

@article{wang2020deepcomplexmri,
  title={DeepcomplexMRI: Exploiting deep residual network for fast parallel MR imaging with complex convolution},
  author={Wang, Shanshan and Cheng, Huitao and Ying, Leslie and Xiao, Taohui and Ke, Ziwen and Zheng, Hairong and Liang, Dong},
  journal={Magnetic resonance imaging},
  volume={68},
  pages={136--147},
  year={2020},
  publisher={Elsevier}
}

@article{neal1992bayesian,
  title={Bayesian learning via stochastic dynamics},
  author={Neal, Radford},
  journal={Advances in Neural Information Processing Systems},
  volume={5},
  year={1992}
}

@article{jospin2022hands,
  title={Hands-on {B}ayesian neural networks—A tutorial for deep learning users},
  author={Jospin, Laurent Valentin and Laga, Hamid and Boussaid, Farid and Buntine, Wray and Bennamoun, Mohammed},
  journal={IEEE Computational Intelligence Magazine},
  volume={17},
  number={2},
  pages={29--48},
  year={2022},
  publisher={IEEE}
}

@article{cole2021analysis,
  title={Analysis of deep complex-valued convolutional neural networks for {MRI} reconstruction and phase-focused applications},
  author={Cole, Elizabeth and Cheng, Joseph and Pauly, John and Vasanawala, Shreyas},
  journal={Magnetic Resonance in Medicine},
  volume={86},
  number={2},
  pages={1093--1109},
  year={2021},
  publisher={Wiley Online Library}
}

@article{mu2021cv,
  title={{CV-GMTINet}: {GMTI} using a deep complex-valued convolutional neural network for multichannel {SAR-GMTI} system},
  author={Mu, Huilin and Zhang, Yun and Jiang, Yicheng and Ding, Chang},
  journal={IEEE Transactions on Geoscience and Remote Sensing},
  volume={60},
  pages={1--15},
  year={2021},
  publisher={IEEE}
}

@inproceedings{barrachina2021complex,
  title={Complex-valued vs. real-valued neural networks for classification perspectives: An example on non-circular data},
  author={Barrachina, Jose Agustin and Ren, Chenfang and Morisseau, Christele and Vieillard, Gilles and Ovarlez, J-P},
  booktitle={IEEE International Conference on Acoustics, Speech and Signal Processing},
  pages={2990--2994},
  year={2021},
  organization={IEEE}
}

@inproceedings{blundell2015weight,
  title={Weight uncertainty in neural network},
  author={Blundell, Charles and Cornebise, Julien and Kavukcuoglu, Koray and Wierstra, Daan},
  booktitle={International Conference on Machine Learning},
  pages={1613--1622},
  year={2015},
  organization={PMLR}
}

@article{cao2019pixel,
  title={Pixel-wise PolSAR image classification via a novel complex-valued deep fully convolutional network},
  author={Cao, Yice and Wu, Yan and Zhang, Peng and Liang, Wenkai and Li, Ming},
  journal={Remote Sensing},
  volume={11},
  number={22},
  pages={2653},
  year={2019},
  publisher={MDPI}
}

@article{gao2018enhanced,
  title={Enhanced radar imaging using a complex-valued convolutional neural network},
  author={Gao, Jingkun and Deng, Bin and Qin, Yuliang and Wang, Hongqiang and Li, Xiang},
  journal={IEEE Geoscience and Remote Sensing Letters},
  volume={16},
  number={1},
  pages={35--39},
  year={2018},
  publisher={IEEE}
}

@inproceedings{gal2016dropout,
  title={Dropout as a {B}ayesian approximation: Representing model uncertainty in deep learning},
  author={Gal, Yarin and Ghahramani, Zoubin},
  booktitle={International Conference on Machine Learning},
  pages={1050--1059},
  year={2016},
  organization={PMLR}
}

@article{trabelsi2017deep,
  title={Deep complex networks},
  author={Trabelsi, Chiheb and Bilaniuk, Olexa and Zhang, Ying and Serdyuk, Dmitriy and Subramanian, Sandeep and Santos, Joao Felipe and Mehri, Soroush and Rostamzadeh, Negar and Bengio, Yoshua and Pal, Christopher J},
  journal={arXiv preprint arXiv:\allowbreak 1705.09792},
  year={2017}
}

@INPROCEEDINGS{ouabi2020stochastic,
  author={Ouabi, Othmane-Latif and Pribić, Radmila and Olaru, Sorin},
  booktitle={28th European Signal Processing Conference}, 
  title={Stochastic Complex-valued Neural Networks for Radar}, 
  year={2021},
  volume={},
  number={},
  pages={1442-1446},
  doi={10.23919/Eusipco47968.2020.9287425}}

@inproceedings{zhang2024accelerating,
  title={Accelerating {MRI} uncertainty estimation with mask-based {B}ayesian neural network},
  author={Zhang, Zehuan and Genci, Matej and Fan, Hongxiang and Wetscherek, Andreas and Luk, Wayne},
  booktitle={IEEE 35th International Conference on Application-specific Systems, Architectures and Processors},
  pages={107--115},
  year={2024},
  organization={IEEE}
}

@inproceedings{fan2023monte,
  title={When {M}onte-{C}arlo Dropout Meets Multi-Exit: Optimizing {B}ayesian Neural Networks on {FPGA}},
  author={Fan, Hongxiang and Chen, Mark and Castelli, Liam and Que, Zhiqiang and Li, He and Long, Kenneth and Luk, Wayne},
  booktitle={ACM/IEEE Design Automation Conference},
  pages={1--6},
  year={2023},
  organization={IEEE}
}

@article{fan2022fpga,
  title={{FPGA}-based acceleration for {B}ayesian convolutional neural networks},
  author={Fan, Hongxiang and Ferianc, Martin and Que, Zhiqiang and Liu, Shuanglong and Niu, Xinyu and Rodrigues, Miguel RD and Luk, Wayne},
  journal={IEEE Transactions on Computer-Aided Design of Integrated Circuits and Systems},
  volume={41},
  number={12},
  pages={5343--5356},
  year={2022},
  publisher={IEEE}
}

@inproceedings{lee2023exploiting,
  title={Exploiting Inherent Properties of Complex Numbers for Accelerating Complex Valued Neural Networks},
  author={Lee, Hyunwuk and Jang, Hyungjun and Kim, Sungbin and Kim, Sungwoo and Cho, Wonho and Ro, Won Woo},
  booktitle={Proceedings of the 56th Annual IEEE/ACM International Symposium on Microarchitecture},
  pages={1121--1134},
  year={2023}
}

@inproceedings{peng2021binary,
  title={Binary complex neural network acceleration on {FPGA}},
  author={Peng, Hongwu and Zhou, Shanglin and Weitze, Scott and Li, Jiaxin and Islam, Sahidul and Geng, Tong and Li, Ang and Zhang, Wei and Song, Minghu and Xie, Mimi and others},
  booktitle={2021 IEEE 32nd International Conference on Application-specific Systems, Architectures and Processors (ASAP)},
  pages={85--92},
  year={2021},
  organization={IEEE}
}

@article{ahmad2024fpga,
  title={{FPGA} Implementation of Complex-Valued Neural Network for Polar-Represented Image Classification},
  author={Ahmad, Maruf and Zhang, Lei and Chowdhury, Muhammad EH},
  journal={Sensors},
  volume={24},
  number={3},
  pages={897},
  year={2024},
  publisher={MDPI}
}

@article{blei2017variational,
  title={Variational inference: A review for statisticians},
  author={Blei, David M and Kucukelbir, Alp and McAuliffe, Jon D},
  journal={Journal of the American Statistical Association},
  volume={112},
  number={518},
  pages={859--877},
  year={2017},
  publisher={Taylor \& Francis}
}

@inproceedings{popa2017complex,
  title={Complex-valued convolutional neural networks for real-valued image classification},
  author={Popa, C{\u{a}}lin-Adrian},
  booktitle={International Joint Conference on Neural Networks},
  pages={816--822},
  year={2017},
  organization={IEEE}
}

@article{cai2018vibnn,
  title={{VIBNN}: Hardware acceleration of {B}ayesian neural networks},
  author={Cai, Ruizhe and Ren, Ao and Liu, Ning and Ding, Caiwen and Wang, Luhao and Qian, Xuehai and Pedram, Massoud and Wang, Yanzhi},
  journal={ACM SIGPLAN Notices},
  volume={53},
  number={2},
  pages={476--488},
  year={2018},
  publisher={ACM New York, NY, USA}
}

@inproceedings{fan2021high,
  title={High-performance {FPGA}-based accelerator for {B}ayesian neural networks},
  author={Fan, Hongxiang and Ferianc, Martin and Rodrigues, Miguel and Zhou, Hongyu and Niu, Xinyu and Luk, Wayne},
  booktitle={2021 58th ACM/IEEE Design Automation Conference (DAC)},
  pages={1063--1068},
  year={2021},
  organization={IEEE}
}

@inproceedings{ferianc2021optimizing,
  title={Optimizing {B}ayesian recurrent neural networks on an {FPGA}-based accelerator},
  author={Ferianc, Martin and Que, Zhiqiang and Fan, Hongxiang and Luk, Wayne and Rodrigues, Miguel},
  booktitle={2021 International Conference on Field-Pro\-gram\-ma\-ble Technology (ICFPT)},
  pages={1--10},
  year={2021},
  organization={IEEE}
}

@inproceedings{awano2020bynqnet,
  title={{BYNQ}Net: {B}ayesian neural network with quadratic activations for sampling-free uncertainty estimation on {FPGA}},
  author={Awano, Hiromitsu and Hashimoto, Masanori},
  booktitle={2020 Design, Automation \& Test in Europe Conference \& Exhibition (DATE)},
  pages={1402--1407},
  year={2020},
  organization={IEEE}
}

@inproceedings{fujiwara2021asbnn,
  title={{ASBNN}: Acceleration of {B}ayesian convolutional neural networks by algorithm-hardware co-design},
  author={Fujiwara, Yoshiki and Takamaeda-Yamazaki, Shinya},
  booktitle={2021 IEEE 32nd International Conference on Application-specific Systems, Architectures and Processors (ASAP)},
  pages={226--233},
  year={2021},
  organization={IEEE}
}

@inproceedings{li2024classification,
  title={Classification, Regression and Segmentation directly from k-Space in Cardiac MRI},
  author={Li, Ruochen and Pan, Jiazhen and Zhu, Youxiang and Ni, Juncheng and Rueckert, Daniel},
  booktitle={International Workshop on Machine Learning in Medical Imaging},
  pages={31--41},
  year={2024},
  organization={Springer}
}

@article{xiao2022partial,
  title={Partial Fourier reconstruction of complex MR images using complex-valued convolutional neural networks},
  author={Xiao, Linfang and Liu, Yilong and Yi, Zheyuan and Zhao, Yujiao and Xie, Linshan and Cao, Peibei and Leong, Alex TL and Wu, Ed X},
  journal={Magnetic Resonance in Medicine},
  volume={87},
  number={2},
  pages={999--1014},
  year={2022},
  publisher={Wiley Online Library}
}

@inproceedings{guo2017calibration,
  title={On calibration of modern neural networks},
  author={Guo, Chuan and Pleiss, Geoff and Sun, Yu and Weinberger, Kilian Q},
  booktitle={International conference on machine learning},
  pages={1321--1330},
  year={2017},
  organization={PMLR}
}

@article{1571417126193283840,
  title={Gradient-based learning applied to document recognition},
  author={LeCun, Yann and Bottou, L{\'e}on and Bengio, Yoshua and Haffner, Patrick},
  journal={Proceedings of the IEEE},
  volume={86},
  number={11},
  pages={2278--2324},
  year={2002},
  publisher={Ieee}
}

@inproceedings{zhang2024hardware,
  title={Hardware-Aware Neural Dropout Search for Reliable Uncertainty Prediction on {FPGA}},
  author={Zhang, Zehuan and Fan, Hongxiang and Chen, Hao and Dudziak, Lukasz and Luk, Wayne},
  booktitle={Proceedings of the 61st ACM/IEEE Design Automation Conference (DAC)},
  pages={1--6},
  year={2024}
}

\end{document}